\documentclass[12pt]{article}
\usepackage{graphicx}

\begin{document}

\noindent {\small CITUSC/00-048\hfill \hfill hep-th/0008164 }



\begin{center}
{\Large \textbf{Survey of Two-Time Physics}}\footnote{%
This research was partially supported by the US. Department of Energy under
grant number DE-FG03-84ER40168. The paper is based on lectures delivered at
the Workshop on Strings, Branes and M-theory, CIT-USC Center, Los Angeles
(March 2000); Second International School on Field Theory and Gravitation,
Vit\'{o}ria, Brazil (April 2000); Quantization Gauge Theory and Strings,
Moscow (June 2000); Non-perturbative Methods in Field and String Theory,
Copenhagen (June 2000); M-theory and Dualities, Istanbul (June 2000).}

\bigskip


\textbf{Itzhak Bars}

{\vskip0.2cm}

\textbf{CIT-USC Center for Theoretical Physics }

\textbf{and}

\textbf{Department of Physics and Astronomy}

\textbf{University of Southern California }

\textbf{Los Angeles, CA 90089-2535, USA}

{\vskip0.2cm}
\end{center}

Two-time physics (2T) is a general reformulation of one-time physics (1T)
that displays previously unnoticed hidden symmetries in 1T dynamical systems
and establishes previously unknown duality type relations among them. This
may play a role in displaying the symmetries and constructing the dynamics
of little understood systems, such as M-theory. 2T physics describes various
1T dynamical systems as different d-dimensional ``holographic'' views of the
same 2T system in $d+2$ dimensions. The ``holography'' is due to gauge
symmetries that tend to reduce the number of effective dimensions. Different
1T evolutions (i.e. different Hamiltonians) emerge from the same 2T theory
when gauge fixing is done with different embeddings of d dimensions inside
d+2 dimensions. Thus, in the 2T setting, the distinguished 1T which we call
``time'' is a gauge dependent concept. The 2T action has also a global
SO(d,2) symmetry in flat spacetime, or a more general d+2 symmetry in curved
spacetime, under which all dimensions are on an equal footing. This symmetry
is observable in many 1T systems, but it remained unknown until discovered
in the 2T formalism. The symmetry takes various non-linear (hidden) forms in
the 1T systems, and it is realized in the same irreducible unitary
representation (same Casimir eigenvalues) in their quantum Hilbert spaces.
2T physics has mainly been developed in the context of particles, including
spin and supersymmetry, but some advances have also been made with strings
and p-branes, and insights for M-theory have already emerged. In the case of
particles, there exists a general worldline formulation with background
fields, as well as a field theory formulation, both described in terms of
fields that depend on d+2 coordinates. All 1T particle interactions with
Yang-Mills, gravitational and other fields are included in the d+2
reformulation. In particular, the Standard Model of particle physics can be
regarded as a gauge fixed form of a 2T theory in 4+2 dimensions. These facts
already provide evidence for a new type of higher dimensional unification.%
\newline
\newpage

\section{Introduction}

Although two-time physics \cite{conf}-\cite{2Tfield} is currently best
understood in simple everyday physics, it originally developed from hints
about two-timelike dimensions in the mathematical structure of M-theory. In
particular the first hint came from the 11-dimensional extended
superalgebra, including the 2-brane and 5-brane charges which has the form 
\cite{townsend} 
\begin{equation}
\left\{ Q_{\alpha },Q_{\beta }\right\} =\gamma _{\alpha \beta }^{\mu }P_{\mu
}+\gamma _{\alpha \beta }^{\mu \nu }Z_{\mu \nu }+\gamma _{\alpha \beta
}^{\mu _{1}\cdots \mu _{5}}Z_{\mu _{1}\cdots \mu _{5}}\,.  \label{M-algebra}
\end{equation}
It was noted that this structure provides a model independent signal for
12-dimensions in M-theory with (10,2) signature \cite{ibjapan}, since the 32
supercharges may be viewed as a Weyl spinor in 12-dimensions and the 528
bosonic charges may be viewed as a 2-form plus a self dual 6-form in
12-dimensions, 
\begin{equation}
P_{\mu }\oplus Z_{\mu \nu }=Z_{MN},\qquad Z_{\mu _{1}\cdots \mu
_{5}}=Z_{M_{1}\cdots M_{6}}^{+}.
\end{equation}
This observation has been generalized in several directions, including
S-theory \cite{stheory} and several of its applications that lend further
support to this view. Taking into account various dualities, 13 dimensions
with (11,2) signature appeared more appealing because in that framework
S-theory can unify type-IIA and type-IIB supersymmetric systems in 10
dimensions.

There are several other observations that support two timelike dimensions.
These include the brane scan \cite{duff}, N=2 superstrings \cite{vafan2},
F-theory \cite{vafa}, U-theory \cite{utheory}, the hints for a 12D super
Yang-Mills or supergravity theory \cite{sezgin}, the AdS-CFT correspondence 
\cite{maldacena}, etc.

The question is whether these hints imply two timelike dimensions exist? Can
they be made manifest in a formulation of the fundamental theory including
explicitly two timelike dimensions with the associated symmetries?
Historically, in previous failed attempts for more timelike dimensions, some
formidable obstacles to overcome included causality and unitarity, the
latter due to ghosts (negative norm states) created by extra timelike
dimensions. Because of these fundamental problems extra timelike dimensions
could not be hidden away by treating them naively like extra spacelike
dimensions and pretending that they are compactified in little circles.

The answer to the fundamental problems could only be a new gauge symmetry
that removes the ghosts and establishes both unitarity and causality.
Two-time physics introduced a new symplectic gauge symmetry which indeed
removes all ghosts, establishes unitarity and causality, and plays a role
analogous to duality \footnote{%
It may be significant that the S,T,U dualities in M-theory or the
Seiberg-Witten dualities in super Yang-Mills theory are also (discrete)
gauge symmetries of a symplectic nature that mix canonically conjugate
quantities (such as electric-magnetic p-brane charges, or windings in space
and Kaluza-Klein momenta, etc).}. Two-time physics is verified in everyday
physics that has already been understood. Besides bringing the new $d+2$
dimensional insights to well known physics, it is hoped that the formalism
would be helpful in the formulation of fundamental physics that remains to
be understood.

The gauge symmetry that plays a fundamental role in two-time physics
historically evolved from attempts to give a dynamical description of
S-theory. In that vein, new gauge symmetry concepts were developed in \cite
{multi} for constrained multi-particle or particle-string systems which
require two-timelike dimensions for a consistent formulation. Eventually
these efforts led to the formulation of the gauge symmetry for a single
particle as given in \cite{conf}.

The gauge symmetry is very natural and could have been explored independent
of M-theory, S-theory or other motivations. It arises as follows: in the
first order formulation of any theory the action has the form $S=\int [\dot{X%
}\cdot P-H(X,P)]$. Up to an irrelevant total derivative the first term can
be rewritten as $\int {\frac{1}{2}}[\dot{X}\cdot P-\dot{P}\cdot X]=$ $\int {%
\frac{1}{2}}\dot{X_{i}}\cdot X_{j}\,\epsilon ^{ij}$, where $%
(X^{M},P^{M})=X_{i}^{M}$ is the Sp(2,R) doublet in phase space. This shows
that the first term in the action has a \textit{global} Sp(2,R) symmetry.
Furthermore, the same Sp(2,R) appears as an automorphism symmetry in the
quantum relations $[X_{i}^{M},X_{j}^{N}]=i\epsilon _{ij}\ \eta ^{MN}$. This
global symmetry is present in every quantum relation of generalized
coordinates and in every first term of the action for any dynamical system.
A fundamental question is: what are the systems for which there is a local
Sp(2,R) symmetry, not only in the first term, but in the full action? Also ,
if such a system exists, what are some interesting generalizations of this
local symmetry? This fundamental question may be taken as the basic starting
point for two-time physics. The two times and all other consequences
(unification of 1T systems, etc.) follow directly from the gauge symmetry.
In particular, two times (no less and no more!) is an outcome of the gauge
symmetry, it is not put in by hand. Thus, two time physics arises from a
gauge principle.

In these lecture notes I will mainly emphasize the concepts of 2T physics. I
will briefly summarize some facts and results whose details are found in the
literature \cite{conf}-\cite{2Tfield}\footnote{%
After much work was done in 2T-physics, including background fields and
field theory, I became aware of some results that were independently
obtained by other authors \cite{Dirac}-\cite{vasilev}. The bulk of this
literature originates with Dirac's formulation in 1936 of a 6-dimensional
field theory formalism whose goal was to understand the conformal SO$\left(
4,2\right) $ symmetry in 4-dimensional spacetime in a linear realization.
Dirac and his followers were not aware of the underlying local Sp$\left(
2,R\right) $ symmetry which plays the fundamental role in 2T physics. This
local symmetry explains the origin of Dirac's field equations in
6-dimensions which he arrived at with a very different reasoning.
Furthermore, one can understand that Dirac's (and other author's) path from
6 to 4 dimensions corresponds to one of the many possible gauge choices in
2T-physics. There was no awareness of the many paths of coming down from $%
d+2 $ to $d$ dimensions which is related to the ability of making a variety
of Sp$(2,R)$ gauge choices. As a result, the older work missed one of the
most important aspects of 2T-physics - namely the ``holographic''
unification of different 1T physical systems (different Hamiltonians) in a
single 2T-action. It also missed the related different realizations of SO$%
\left( d,2\right) $ that have different interpretations than conformal
symmetry, thus reflecting the presence of the underlying $d+2$ dimensions.
It was fortunate to have been unaware of this literature while 2T physics
developed, for it could have derailed the exploration of the new aspects
revealed by 2T-physics. Of course, the past work did not address the $d+2$
formulation of more modern topics such as supersymmetry, p-branes, etc. that
have been developed in the context of 2T-physics.}.

\section{2T follows from gauge symmetry}

In the worldline formulation of particles the new gauge symmetry, and its
generalizations, act on phase space. The first class constraints associated
with the gauge symmetry insure unitarity (no ghosts). These constraints have
non-trivial solutions only if the \textit{target spacetime} has two timelike
dimensions. Thus, the gauge symmetry demands that neither fewer nor more
timelike dimensions are permitted in the description of a single point
particle (the many body theory described by the corresponding field theory
also must have exactly two timelike dimensions).

For spinless particles the gauge symmetry is the Sp$(2,R)$ that acts on
position-momentum ($X^{M},P^{M}$) as a doublet. For spinning particles the
gauge symmetry is OSp$(n|2)$ which acts on super phase space ($\psi
_{a}^{M},X^{M},P^{M}$) in its fundamental representation (where $a=1,\cdots
,n$). With spacetime supersymmetry the gauge symmetry is enlarged with a new
version of kappa supersymmetry in $d+2$ dimensions.

One consequence of the gauge symmetry in all cases is the requirement of two
timelike dimensions in target spacetime since otherwise the constraints have
no non-trivial physical solutions. Thus the 2T formulation is dictated by
the gauge symmetry.

For example for spinless particles the Sp($2,R$) constraints are $%
X^{2}=P^{2}=X\cdot P=0$. If the target spacetime signature in these dot
products is Euclidean the only solution is $X^{M}=P^{M}=0$, and the theory
is trivial. If it has one timelike direction, then $X^{M}$ and $P^{M}$ can
only be lightlike vectors that are parallel, so there is no angular
momentum, which is also trivial. If the signature has three timelike
directions, then there are too many ghosts that cannot be removed by the Sp($%
2,R$) gauge symmetry. Thus, only for the case with two timelike dimensions
the constraints have non-trivial solutions and the gauge symmetry removes
all ghosts. In particular, out of the three gauge parameters in Sp($2,R$),
one is related to the familiar $\tau $-reparametrizations, while the other
two have signatures such that they can remove one spacelike and one timelike
dimensions \footnote{%
If one considers gauging Sp(2n), instead of Sp(2), then one finds that the
system describes n particles with 2n times in $(d+2n-2,2n)$ dimensions \cite
{multisp2n}. After partial gauge fixing and solving a subset of the Sp(2n)
constraints such a system reduces to the n-time n-particle theories in $%
(d+n-2,n)$ dimensions discussed in \cite{multi}. Fully solving all
constraints one obtains $n$ particles each in $(d-1,1)$ dimensions, but each
using a different timelike coordinate as embedded in the higher dimensions.
So, to describe a single particle (and the associated field theory, i.e. 1T
many-body system) we can only have local Sp(2) and only two times.}.

For all other cases of gauge (super)symmetries mentioned above - or their
generalizations to strings and branes - their content is as follows: there
is just enough gauge symmetry to remove one timelike and one spacelike
directions from all SO$(d,2)$ vectors and also remove half of the SO($d,2$)
spinor components to arrive at SO$(d-1,1)$ spinors. Further generalizations
of the gauge symmetries (see e.g. toy M-model in section 7.3) also remove
ghost components from SO$(d,2)$ tensors (such as antisymmetric tensors
associated with branes).

\section{Unification of 1T systems, ``holography''}

The simplest action with an Sp$(2,R)$ gauge symmetry is \cite{conf} 
\begin{equation}
L=\dot{X}_{1}\cdot X_{2}-\frac{1}{2}A^{ij}X_{i}\cdot X_{j}  \label{L}
\end{equation}
where $X_{i}^{M}=\left( X^{M},P^{M}\right) $ is the Sp$\left( 2,R\right) $
doublet, and $A^{ij}$ is the Sp$\left( 2,R\right) $ gauge potential. The
equation of motion of the gauge potential leads to the constraints, hence
the target spacetime must have signature $(d,2)$ as an outcome of the Sp$%
\left( 2,R\right) $ gauge symmetry.

This system can be embellished with the addition of background fields \cite
{emgrav} (Yang-Mills, gravity, others), spinning degrees of freedom \cite
{spin}, spacetime supersymmetry \cite{super2t}-\cite{toyM}, and can be
generalized to strings and p-branes \cite{string2t}. For each generalization
the Sp(2,R) gauge symmetry is enriched with a modified gauge symmetry.

A given 2T-physics system defined by an action in $(d,2)$ dimensions $X^{M}$%
, displays a unification of a class of 1-time dynamical systems in $(d-1,1)$
dimensions $x^{\mu }$. That is, a given 2T action modulo gauge symmetries,
is equivalent to many 1T actions with different 1T dynamics. The extra
1-spacelike and 1-timelike dimensions (1,1) are interchangeable with gauge
degrees of freedom. The different embeddings of $d$ dimensions $x^{\mu }$
inside $d+2$ dimensions $X^{M}$ produce different dynamics (different
Hamiltonians) in the chosen $d$ dimensions. The dynamical systems obtained
in this way belong to a class defined by a given 2T action. Changing the 2T
action (e.g. changing background fields, etc.) changes the class. The gauge
symmetry relates the different $d$ dimensional 1T actions (in the same
class) to each other in a way reminiscent of ``duality''.

One may say that various 1T actions in $d$ dimensions are different
``holographic'' descriptions of the same 2T system in $d+2$ dimensions.
There is an equivalence between a family of different ``dynamics'' and a
family of different ``holographic'' views. The simplest and most symmetric
view is the non-holographic $d+2$ dimensional description.

The essential ideas of $d+2$ dimensional unification can already be
understood in the simplest case. For the action (\ref{L}) some members of
the class of 1T spinless particle systems is given in Fig.1. The action (or
equations of motion) of those 1T dynamical systems emerge by gauge fixing
this 2T Lagrangian (or its equations of motion).


\begin{figure}[tbp]
\centering
\includegraphics[height=10cm]{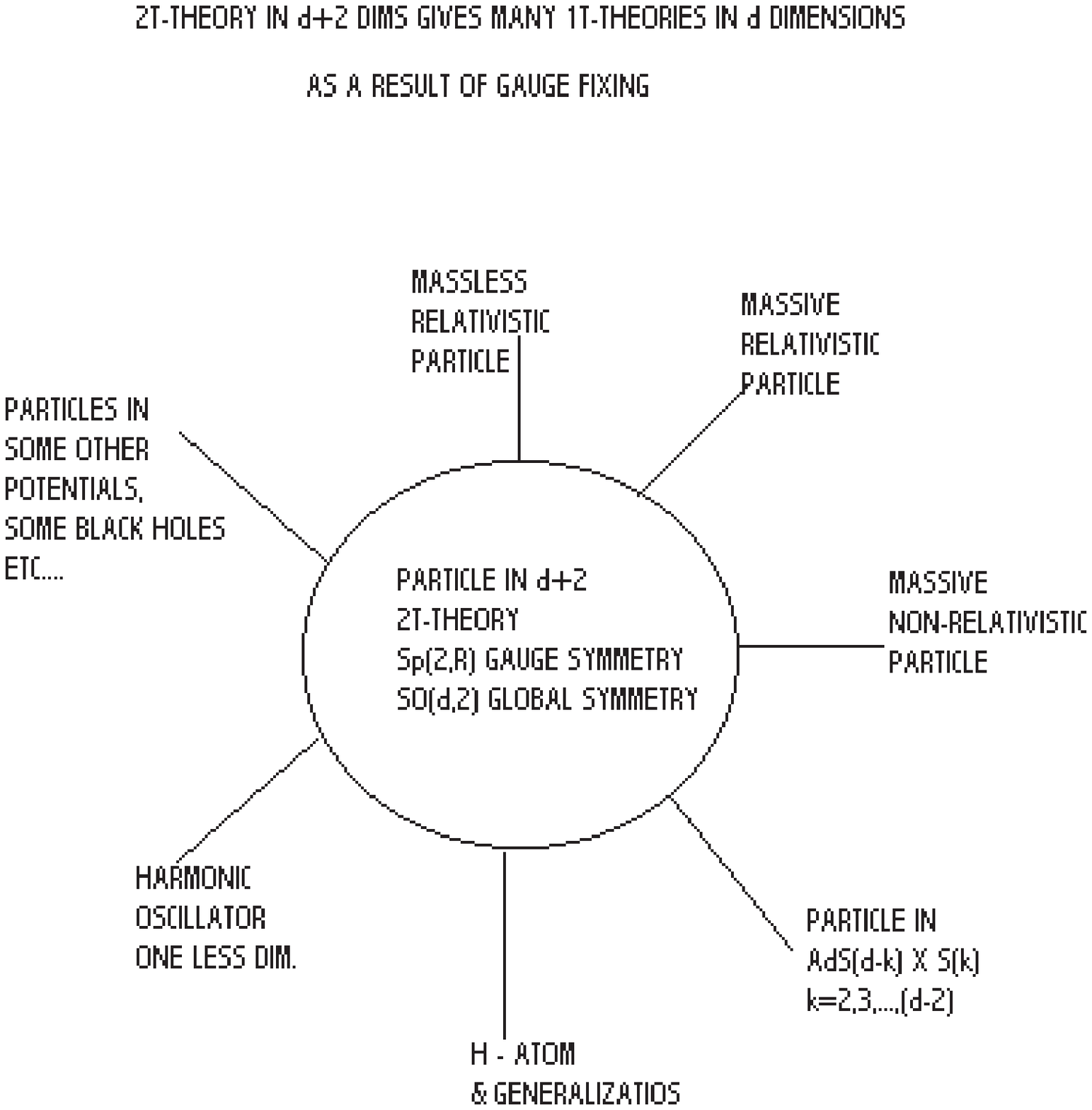}
\caption{{}}
\end{figure}


\begin{itemize}
\item  {Massless relativistic particle: $X^{+^{\prime }}(\tau )=1,\ \
P^{+^{\prime }}(\tau )=0$.}

\item  {Massive non-relativistic particle: $P^{+^{\prime }}(\tau )=m,\ \
P^0(\tau )=0$.}

\item  {Massive relativistic particle: $P^{+^{\prime }}(\tau )=m,\ \
P^0(\tau )=0$.}

\item  {Particle on AdS$_d$: $X^{d-1}(\tau )=1,\ \ P^{+^{\prime }}(\tau )=0$,}
\end{itemize}
and so on for the H-atom, harmonic oscillator, etc. noted in the figure, and
more (see \cite{conf}\cite{lifting} for details). In each case one timelike
dimension and one spacelike dimension is eliminated from each vector $%
(X^{M},P^{M})$ by the gauge choice and the solution of the two constraints $%
X^{2}=X\cdot P=0$. The remaining gauge choice and constraint $P^{2}=0$ is
associated with the $\tau $-reparametrization gauge symmetry, and it may be
convenient to gauge fix it to complete the relation between $\tau $ and the
1T-dimension embedded in $\left( d,2\right) $ target spacetime $X^{M}\left(
\tau \right) $. The theory, as expressed by the remaining coordinates, is
then forced to evolve as a function of the remaining 1T timelike dimension.
The Hamiltonian (canonical conjugate to the 1T dimension) that describes
this time evolution takes different forms in each case in terms of the
remaining canonical degrees of freedom. Such embeddings of ``time'' as a
curve inside $d+2$ dimensions can be done in an infinite number of ways,
thereby producing a class of related 1T dynamical systems from the same 2T
theory.

When the 2T Lagrangian is changed by including spin, supersymmetry,
background fields (but always maintaining a local Sp$\left( 2\right) $
symmetry), the class of related 1T systems changes accordingly. Thus, the
class of unified 1T systems is defined by the 2T action.

It is a fact, by construction, that the 1T ``time" and its associated
Hamiltonian are gauge dependent concepts in this setting , and all 1T
systems in the same class are gauge related to the same 2T theory. They can
all be gauge transformed to each other, and these gauge transformations are
the analogs of ``duality". These are concepts that physicists will take some
time to get accustomed to, but they will do so with a little effort, because
these facts are easily established in simple systems as well as generally.

The relations between 1T theories become more understandable by
concentrating on gauge invariants such as the SO$\left( d,2\right) $ \textit{%
global} symmetry generators 
\begin{equation}
L^{MN}=X_{i}^{M}X_{j}^{N}\epsilon ^{ij}=X^{M}P^{N}-X^{N}P^{M}.
\label{Lorentz}
\end{equation}
In any 1T system the $d$ canonical coordinates that remain after gauge
fixing $\left( x^{\mu },p^{\mu }\right) $ can be rewritten as functions of
the gauge invariant $L^{MN}$. Therefore in any fixed 1T system all physical
quantities $F\left( x,p\right) $ can be rewritten in terms of functions of
the gauge invariant operators $f(L^{MN})$. Those same functions have a
different expression in terms of the canonical variables in another fixed 1T
system $\tilde{F}\left( \tilde{x}^{\mu },\tilde{p}^{\mu }\right) $. But
because of the gauge invariance of $f(L^{MN})$ they are guaranteed to
produce identical results$\ F\left( x,p\right) =f(L^{MN})=\tilde{F}\left( 
\tilde{x}^{\mu },\tilde{p}^{\mu }\right) $. This observation permits the
establishment of many relations between the 1T systems in the same class. As
long as one computes the same functions of $f(L^{MN})$ it does not matter in
which 1T system it is computed. This is the test of the dualities which
establishes that indeed all of the 1T systems in the same class are related
to the same 2T theory\footnote{%
Some intuition for what is going on may be helpful for readers that are
familiar with canonical transformations, by noting that there is a relation
between the local Sp(2,R) ``duality'' transformations from one 1T system to
another 1T system, and generalized canonical transformations that connect
the same systems including transformations of the time coordinate and the
Hamiltonian. While this observation may be helpful initially to digest the
new insights for simple systems, this relation by itself is not helpful more
generally in understanding other phenomena in 2T physics (e.g. the global
symmetry discussed in the next paragraph). Furhtermore, generalized
canonical transformations is not a well developed method in the context of
spin, supersymmetry (with generalized local kappa), field theory, p-branes,
M-theory etc.. Therefore, the gauge symmetry formalism of 2T physics is a
far more superior and useful language even for the simple cases, and is a
more general method in the long run.}. All this is easily carried out in the
classical theory. So one may use a convenient gauge fixed 1T version to
perform computations that apply in all other related 1T systems (for example
solving equations of motion, etc.). In the quantum theory, operator ordering
in non-linear functions can produce anomalies, and one needs to find the
correct ordering in transforming from one 1T system to another (see \cite
{conf}-\cite{spin} for some examples of quantum ordering, in particular in
the computation of the Casimir operator $\frac{1}{2}L^{MN}L_{MN}$ in
different gauge fixed systems).

\section{Global symmetry, one evidence for $d+2$ dimensions}

In flat spacetime, before gauge fixing, the 2T action in (\ref{L}) has a
global SO$\left( d,2\right) $ symmetry which is linearly realized on $d+2$
coordinates $X^{M}$ and is manifest in the action. This is the Lorentz
symmetry in $d+2$ dimensions which treats all coordinates on an equal
footing. This symmetry is gauge invariant (commutes with the Sp$\left(
2,R\right) $ gauge transformations) so its generators (\ref{Lorentz}) are
physical observables. By the gauge invariance of the action (\ref{L}) and of
the symmetry generators (\ref{Lorentz}) the same symmetry is present in
every 1T action derived from the same 2T action by gauge fixing. In each 1T
action (such as those that correspond to Fig.1) this symmetry is
non-linearly realized in different ways on the fewer $d$ coordinates $x^{\mu
}$ singled out by a fixed gauge.

When the system is in interaction with background fields (see next section),
the global SO$\left( d,2\right) $ symmetry is modified to the Killing
symmetries permitted by those fields. In the field theoretic formulation
(see below), in the presence of gravity, the SO$\left( d,2\right) $ symmetry
is elevated to general coordinate invariance and Yang-Mills type gauge
symmetries in $d+2$ dimensions. Whatever the target spacetime symmetry may
be, it is also present (perhaps in a hidden form) in the 1T systems derived
from the 2T action. This simply follows from the reasoning in the previous
paragraph.

This reasoning in the 2T physics formulation permitted for the first time
the discovery of SO$(d,2)$ symmetry in a variety of familiar systems in
``everyday physics'' \cite{conf}-\cite{spin}. In one 1T gauge, corresponding
to the massless particle in Fig.1, the hidden symmetry was very familiar as
the conformal SO$\left( d,2\right) $ symmetry in massless systems. In
another 1T gauge, corresponding to the H-atom in Fig.1, the hidden symmetry
was also understood a long time ago, as a dynamical symmetry of the usual
H-atom which describes all of its quantum levels as a single irreducible
representation of SO(4,2) (although no-one had noticed before that it is the
classical symmetry of the H-atom action). For all other 1T gauge choices in
Fig.1 the SO$(d,2)$ symmetry of the classical action has been verified \cite
{conf}\cite{lifting}\cite{spin}. Of course, the general argument guarantees
that the symmetry is present at the classical level in every imaginable
gauge choice of 1T. However, except for the two cases mentioned above, for
all other cases the presence of the SO$\left( d,2\right) $ symmetry came as
a surprise to physicists.

It is now possible to claim that the various forms of SO$\left( d,2\right) $
symmetry, such as conformal symmetry or other forms, are nothing but an
expression of the higher dimensional $\left( d+2\right) $ nature of the
system, since the symmetry is nothing but the same Lorentz symmetry of the
unified 2T action that treats all dimensions on the same footing. The gauge
choice blurs the higher dimensional nature of the system, but the symmetry
is there to assert the presence of the higher dimensions. Therefore, there
is a sense in which there are $d+2$ dimensions even when we concentrate on
the 1T description of the system. A similar argument applies when SO$\left(
d,2\right) $ is replaced by a Killing symmetry in the presence of background
fields.

Thus, the establishment of the \textit{target spacetime} hidden SO$\left(
d,2\right) $ or similar Killing symmetry in many familiar 1T actions at the
classical level is one evidence for the presence and relevance of the higher
dimensions. The extra 1-timelike and 1-spacelike dimensions appear to have a
different physical role than the remaining $d$ dimensions in a given fixed
gauge from the ``holographic'' point of view of 1T physics. But one can
claim that this is only one ``non-democratic'' way of observing the full
system. In a different gauge, a different set of ``holographic'' $d$%
-dimensions get distinguished to describe the same $d+2$ dimensional system.
So there is a point of view in which the $(1,1)$ gauge degrees of freedom
and the $\left( d-1,1\right) $ \ dimensions distinguished by a gauge choice
are really on the same footing as far as the SO$\left( d,2\right) $ symmetry
or the spacetime dimensionality of the system is concerned.

What happens at the quantum level? To express the symmetry correctly, it is
necessary to order carefully the operators $\left( x^{\mu },p^{\mu }\right) $
that appear in non-linear expressions of the generators $L^{MN}$. This has
been done successfully in a few of the cases given in Fig.1, and then it was
possible to show that the SO$\left( d,2\right) $ symmetry is realized in a
Hilbert space which corresponds to the \textit{same unitary representation}
of SO$\left( d,2\right) $ (same Casimir eigenvalues) for 1T systems derived
from the same 2T action \cite{conf}-\cite{spin}. The various 1T quantum
systems are distinguished from each other by diagonalizing different subsets
of compatible operators (functions of SO$(d,2)$ generators) including the
distinguished 1T Hamiltonian, but always staying in the Hilbert space with
the same eigenvalues of the SO$(d,2)$ Casimir operators, thus the same
unitary representation of SO$\left( d,2\right) $. The transformation from a
1T system to another one involves the unitary transformation from one set of
compatible operators to another one, but this transformation (which is
related to the Sp$\left( 2,R\right) $ gauge symmetry) commutes with the SO$%
\left( d,2\right) $ Casimirs and hence does not change the SO$\left(
d,2\right) $ representation, it only changes the basis. In this quantum
space the unification of the 1T systems and their duality-like relationships
(of the type mentioned in the previous section $F\left( x,p\right) =\tilde{F}%
\left( \tilde{x}^{\mu },\tilde{p}^{\mu }\right) =f(L^{MN})$) could be
computed most directly by using group theoretical techniques\footnote{%
Generally quantum operator ordering may not be easy for generic gauge
choices. Normally the system in a given gauge would not be correctly
quantized until the symmetry generators close and have the same Casimir
eigenvalues as any other gauge (see \cite{conf}-\cite{spin} for explicit
cases). However, one may raise the question if there are gauges in which the
correct ordering could never be resolved, thus having fundamental anomalies
in that gauge. It is not presently known if this may be the case for some 2T
actions, in particular with background fields. In any case, anomalies of 
\textit{global symmetries} (as opposed to local symmetries) would not
invalidate the 2T theory; they would simply be a property of the theory
similar to anomalies in global symmetries encountered in field theory (e.g.
Adler-Bardeen anomaly, conformal anomaly).}.

\section{Background Fields}

The simplest action without spin in (\ref{L}) is generalized to 
\begin{equation}
S=\int d\tau (\partial _{\tau }X^{M}P_{M}-\frac{1}{2}A^{ij}Q_{ij}\left(
X,P\right) \,),  \label{Lbackgr}
\end{equation}
where a more general function of phase space $Q_{ij}\left( X,P\right) $
replaces the previous expressions $Q_{11}=X^{2},$ $Q_{12}=X\cdot P$ and $%
Q_{22}=P^{2}.$ It is shown in \cite{emgrav} that this action has local Sp$%
\left( 2,R\right) $ gauge symmetry provided $A^{ij}$ transforms like a
standard gauge field and $Q_{ij}\left( X,P\right) $ \ is any function of
phase space that satisfies the Sp$\left( 2,R\right) $ Lie algebra under
Poisson brackets. From the equations of motion of $A^{ij}$ we deduce that
the constraints $Q_{ij}\left( X,P\right) =0$ define the physical states.

A general approach for finding non-trivial models of $Q_{ij}\left(
X,P\right) $ is to setup a series expansion in powers of $P^{M}$ for each $%
Q_{ij}$ with coefficients that are arbitrary functions of $X^{M}.$ These
coefficients are background fields $\phi _{M_{1}\cdots M_{k}}\left( X\right) 
$ with indices that are contracted with the powers of $P^{M_{i}}.$ Imposing
the Sp$\left( 2,R\right) $ algebra under Poisson brackets requires the
background fields to satisfy certain equations. The solution of these
equations determines partially the form of the background fields, but still
leaves undetermined functions that can be chosen according to physical
considerations. As shown in \cite{emgrav}, by keeping the lowest possible
powers of $P,$ one finds that all known fundamental interactions of
particles (Yang-Mills, gravity, others) can be formulated generally in the
2T formalism. The form of the $Q_{ij}$ are then 
\begin{eqnarray}
Q_{11} &=&V^{M}V_{M}+\cdots ,\quad Q_{12}=V^{M}\left( P_{M}+A_{M}\right)
+\cdots \\
Q_{22} &=&U+G^{MN}\left( P_{M}+A_{M}\right) \left( P_{N}+A_{N}\right) +\cdots
\end{eqnarray}
where $A_{M}\left( X\right) ,$ $G_{MN}\left( X\right) ,$ $U\left( X\right) $
are the background fields for Yang-Mills, gravity and a scalar potential in $%
d+2$ dimensions. The dots $\left( \cdots \right) $ represent the terms with
higher powers of $P$ including coefficients that represent higher spin
fields. When all higher spin fields vanish one can show that the field $%
V^{M}\left( X\right) $ satisfies the coupled equations 
\begin{equation}
V_{M}=\frac{1}{2}\partial _{M}\left( V^{K}V_{K}\right) =G_{MN}V^{N},\quad
V^{M}F_{MN}=0,\quad V^{M}\partial _{M}U=-2U,\quad \pounds _{V}G_{MN}=2G_{MN},
\label{spinless}
\end{equation}
where $\pounds _{V}$ is the Lie derivative with respect to $V.$ These
equations are generally covariant and Yang-Mills gauge invariant in $d+2$
dimensions.

After solving these equations explicitly, and imposing the constraints $%
Q_{11}=Q_{12}=0$ in a particular 1T gauge (going from $d+2$ to $d$ in a
gauge similar to the free massless particle case of Fig.1), it was shown in 
\cite{emgrav} that the $d+2$ equations of motion and the action (\ref
{Lbackgr}) collapse to the system 
\begin{equation}
L=\frac{1}{2A^{22}}\dot{x}^{\mu }\dot{x}^{\nu }g_{\mu \nu }\left( x\right) -%
\frac{A^{22}}{2}u\left( x\right) -\dot{x}^{\mu }A_{\mu }\left( x\right)
\label{interaction} 
\end{equation}
which describes the motion of a relativistic spinless particle interacting
with the arbitrary backgrounds $A_{\mu },g_{\mu \nu },u$ representing any
gravitational, Yang-Mills or other interactions in $d$ dimensions.

The analysis may be repeated for other 1T gauges (such as those related to
Fig.1) to establish relationships among interacting systems and unify them
as a 2T theory in a given background described by the action (\ref{Lbackgr})
in $d+2$ dimensions.

The analysis in \cite{emgrav} also included spinning particles by using OSp$%
\left( n|2\right) $ instead of Sp$\left( 2,R\right) $. In this case there is
a spin connection $\omega _{M}^{ab}\left( X\right) $ and a vector $%
V^{a}\left( X\right) $ where the index $a$ is a tangent space index in $d+2$
dimensions. Then one finds that the OSp$\left( n|2\right) $ gauge symmetry
of the action requires that the vector $V^{M},$ soldering form $E_{M}^{a}$, metric $%
G_{MN}$, torsion $T_{MN}^{a}=D_{[M}E_{N]}^{a},$ curvature $R=d\omega +\omega
^{2},$ are all constructed purely from $\omega _{M}^{ab}\left( X\right) $
and $V^{a}\left( X\right) $%
\begin{equation}
E_{M}^{a}=D_{M}V^{a},\quad V^{M}=E_{a}^{M}V^{a},\quad
G_{MN}=E_{M}^{a}E_{N}^{b}\eta _{ab},\quad T_{MN}^{a}=R_{MN}^{ab}V_{b},
\label{definitions}
\end{equation}
where the covariant derivative $D_{M}$ uses the spin connection. The only
restriction on $V^{a}$ and $\omega _{M}^{ab}$ \ is similar to the one
satisfied by the Yang-Mills field 
\begin{equation}
V^{M}R_{MN}^{ab}=0,\quad V^{M}F_{MN}=0.  \label{spinning}
\end{equation}
All the equations in (\ref{spinless}) are solved by the definitions in (\ref
{definitions}) and the solutions\footnote{%
Up to a general coordinate ransformation one can choose a coordinate basis
such that $X^{M}=\left( \kappa ,\rho ,x^{\mu }\right) $ while $V^{M}=\left(
\kappa ,\rho ,0\right) .$ Then, up to a local SO$\left( d,2\right) $ tangent
space gauge transformation, and a Yang-Mills gauge transformation, the
solution takes the following form 
\begin{equation}
V^{a}=\kappa v^{a},\quad \omega _{M}^{ab}=\left( \frac{1}{\kappa }u^{ab},-%
\frac{1}{\rho }u^{ab},\omega _{\mu }^{ab}\right) ,\quad A_{M}=\left( \frac{1%
}{\kappa \,}a,-\frac{1}{\rho \,}a,A_{\mu }\right)
\end{equation}
where the functions $v^{a},u^{ab},\omega _{\mu }^{ab},a,A_{\mu }$ are all
arbitrary functions of $x^{\mu }$ and the ratio $\rho /\kappa ,$ while $%
E_{M}^{a}=\left( \left( v^{a}+w^{a}\right) ,-\frac{\kappa }{\rho }%
w^{a},\kappa e_{\mu }^{a}\right) $ with $e_{\mu }^{a}=D_{\mu }v^{a}$ and $%
w^{a}=\kappa \partial _{\kappa }v^{a}+u^{ab}v_{b}.$ The $R_{\kappa \rho
}^{ab}$ and $T_{\kappa \rho }^{a}$ components of curvature and torsion
vanish in this coordinate basis, but the remaining components are generally
non-zero. The general solution is valid before considering the constraints.
The constraints $Q_{11}=Q_{12}=0$ are uniquely satisfied by setting $\rho
=0=p_{\kappa }.$ Then $e_{\mu }^{a}\left( x\right) $ together with $A_{\mu
}\left( x\right) $ describe arbitrary gravitational and gauge interactions
in the remaining $d$ dimensions $x^{\mu }$, including torsion$.$} of (\ref
{spinning}).

Thus, all fundamental interactions of particles for any spin in $d$
dimensions have a fully
generally covariant and Yang-Mills gauge invariant description in $d+2$
dimensions. As in the background free case, many dynamical systems of an
interacting particle in $d$ dimensions can be unified as a single 2T theory
in $d+2$ dimensions (different gauge choices to embed $d$ in $d+2$). The
resulting classes of 1T systems (similar to Fig.1, but with background
fields) have duality-type relations among them reflecting their common
origin and hidden $d+2$ dimensional symmetries.

\section{2T Field Theory, unified 1T field theories}

1T field theory follows from imposing worldline theory constraints on the
physical quantum states. Thus the constraint $p^{2}=0$ applied on states
gives the Klein-Gordon equation $\partial _{\mu }\partial ^{\mu} \phi =0,$ 
which follows from the
free field Lagrangian $L=\frac{1}{2}\left( \partial _{\mu }\phi \right)
^{2}. $ Field interactions may then be added to consider an interacting
field theory. We showed that one can derive many 1T worldline theories from
the same 2T worldline theory by choosing gauges and solving explicitly the
two constraints $X^{2}=X\cdot P=0.$ The remaining constraint $P^{2}=0$ can
take many different forms in terms of the remaining $d$ dimensions. For the
relativistic massless particle in Fig.1 it is $P^{2}=p^{2}=0,$ while for the
massive non-relativistic particle in Fig.1 it is $P^{2}=-2mH+\mathbf{p}%
^{2}=0,$ etc. Imposing $H=\mathbf{p}^{2}/2m$ on the states gives the
Schr\"{o}dinger equation $i\partial _{t}\psi =-\frac{1}{2m}\nabla ^{2}\psi $
that follows from the action $L=i\psi ^{\ast }\partial _{t}\psi -\frac{1}{2m}%
\mathbf{\nabla }\psi ^{\ast }\cdot \mathbf{\nabla }\psi .$ Thus, various
1T field theories follow from the same 2T worldline theory when
quantization is performed in different 1T gauges. We will next show that we
can derive the family of 1T field theories from the same 2T field theory.

We quantize the worldline system covariantly in $d+2$ dimensions 
keeping the SO$\left(d,2\right) $ symmetry manifest. This is done by 
imposing the Sp$\left(2,R\right) $ constraints on the states to obtain 
the set of Sp$\left(2,R\right) $ gauge invariant physical states. 
However, since the constraints
are non-Abelian we can diagonalize simultaneously only the Casimir and one
of the Sp$\left( 2,R\right) $ generators, and then restrict to the singlet
sector. When this procedure is applied \cite{2Tfield} to the spinless
particle system (\ref{L}) the result is the manifestly SO$\left( d,2\right) $
covariant field equations in $d+2$ dimensions 
\begin{equation}
X^{2}\Phi \left( X\right) =0,\quad X^{M}\partial _{M}\Phi \left( X\right) =-%
\frac{d-2}{2}\Phi \left( X\right) ,\quad \partial ^{2}\Phi \left( X\right)
=g\Phi ^{\left( d+2\right) /\left( d-2\right) }.  \label{scalar}
\end{equation}
The last equation includes the only possible interaction consistent with the
Sp$\left( 2,R\right) $ gauge singlet condition. The last equation, which we
call the ``dynamical equation'' is derived from a Lagrangian 
\begin{equation}
L=-\frac{1}{2}\partial ^{M}\Phi \partial _{M}\Phi -g\frac{d+2}{d-2}\Phi
^{2d/\left( d-2\right) }  \label{scalarL}
\end{equation}
while the first two equations which we call the ``kinematic equations'' are
considered as subsidiary conditions which are not derived from this 
Lagrangian. Note that the kinematic equations
correspond to the classical constraints $X^{2},X\cdot P.$ The equations in (%
\ref{scalar}) at $g=0$ are identical to those derived by Dirac in 1936 with
a rather different approach \cite{Dirac}, with no knowledge of the
underlying Sp$(2,R)$ gauge symmetry. Now we see that his equations are
simply the Sp$\left( 2,R\right) $ singlet condition on physical states.

Dirac showed that the 4-dimensional Klein-Gordon equation 
$\partial _{\mu }\partial ^{\mu}phi =0 $ 
follows from these SO$\left( 4,2\right) $ covariant equations after
solving the kinematic equations and coming down from 6 to 4 dimensions. He
thus proved that the conformal symmetry SO$\left( 4,2\right) $ of the
massless Klein-Gordon equation becomes manifest in the 
form of (\ref{scalar}).

However, Dirac and his followers who were not aware of the underlying Sp$%
\left( 2,R\right) $ gauge symmetry, did not notice that it is possible to
come down from 6 to 4 dimensions in very different ways by choosing
different 1T gauges. The form of the 1T field theory depends on the
embedding of $d$ dimensions in $d+2$ dimensions and solving the \textit{%
kinematic} equations by using those coordinates. Replacing the solution in
the last equation in (\ref{scalar}) yields the 1T \textit{dynamical}
equation in $d$ dimensions. Furthermore, the 2T Lagrangian (\ref{scalarL})
reduces to the correct 1T field theory Lagrangian. 
Using this procedure, it was shown in 
\cite{2Tfield} that the Schr\"{o}dinger equation and other equations that
correspond to the first quantized field equations for the systems in Fig.1
follow from (\ref{scalar}). Therefore, the same 2T field theory in $d+2$
dimensions unifies many types of 1T field theories in $d$ dimensions, such
as those that correspond to Fig.1. The local field interactions in $d+2,$ as
in (\ref{scalarL}), result in various forms of interaction in different 1T
gauges. The interaction has the same form as (\ref{scalarL}) in the gauge
that corresponds to the relativistic massless particle in Fig.1, but can be
more involved in other gauges. For examples see \cite{2Tfield}.

A similar approach works for spinning particles. The SO$\left( d,2\right) $
covariant quantization of the 2T worldline theory leads to a formulation of
2T interacting classical field theory for spinning fields. In $d+2$
dimensions spinors have twice as many components as in $d$ dimensions, so
how can the $d+2$ field theory become the same as the $d$ dimensional field
theory? Interestingly, the OSp$\left( 1|2\right) $ gauge singlet condition
produces a new type of spinor field equation (and Lagrangian) which have a
kappa type symmetry that cuts down the number of spinor components by a
factor of 2. The process also generates the Yang-Mills and gravitational
equations with consistent interactions in $d+2$ dimensions. One finds that
the equations in (\ref{spinning}) together with $X^{2}=0$ are the
kinematical equations for the Yang-Mills and gravitational fields, while the
dynamical equations, including interactions, follow from a Lagrangian as
given in \cite{2Tfield}. Then one sees that the Standard model may be
regarded as a 2T field theory in 6-dimensions which is reduced to 4
dimensions in a particular gauge.

Our current understanding of the 2T field theory formulation is somewhat
incomplete. This is because the kinematic equations such as those in (\ref
{scalar},\ref{spinning}) are imposed as additional field equations which do
not follow from the field theory Lagrangian. They correspond to the
worldline theory constraints $Q_{11}=Q_{12}=0$. A more satisfactory
situation would be to derive all the equations, not only the dynamical
equation, directly in the field theory formalism. One approach is to
introduce Lagrange multipliers, but this seems artificial. A more natural 2T
field theory may need to be based on fields that depend on phase space $\phi
\left( X,P\right) $ and which transform under a local Sp$\left( 2,R\right) $
symmetry (this is somewhat reminiscent of non-commutative geometry). Then
the Sp$\left( 2,R\right) $ singlet equations (both kinematic and dynamic)
could follow from the gauge symmetry of the field theory rather than the
gauge symmetry of the worldline theory. This formulation remains as a
challenge.

Although the 2T unification of 1T systems can be examined in either the
worldline or field theory forms, the worldline approach provides a better
understanding of the underlying gauge symmetries at this stage, while the
field theory formulation provides an approach for interactions among fields.

Second quantization of 2T physics may be considered. The improvement of the
field theory mentioned above would probably help in the formulation of the
second quantization.

\section{Local and global 2T Supersymmetry}

The generalization of the 2T theory to spacetime supersymmetry emerged in
several steps by developing a new approach to both global and local kappa
supersymmetries \cite{super2t}-\cite{toyM}. The generalized form of (\ref{L}%
) that applies in several situations of physical interest is 
\begin{equation}
S=\int d\tau \left[ \left( \dot{X}_{1}^{M}X_{2}^{N}-\frac{1}{2}%
A^{ij}X_{i}^{M}X_{j}^{N}\right) \eta _{MN}-Str\left( L\left( \partial _{\tau
}gg^{-1}\right) \right) \right] ,  \label{Lsuper}
\end{equation}
where $g\in G$ is a supergroup element, and $L=L^{MN}\Gamma _{MN}$ is a
coupling of the Cartan form $\partial _{\tau }gg^{-1}$ to the orbital 
SO$\left( d,2\right) $ Lorentz generators 
$L^{MN}=\varepsilon ^{ij}X_{i}^{M}X_{j}^{N}.$ 
The supergroup element\ $g$ contains fermions $\Theta $  that are now
coupled to the $X^M_i$.

Let us define $G_{L}$ as the supergroup under 
\textit{left multiplication }$g\leftarrow g_{L}g$. We require that the
bosonic subgroup of $G_{L}$ contain at least 
SO$\left( d-k,2\right) \times $ SO$\left( k\right) $ 
for some $k=0,\cdots ,d-2.$ This is identified
with part of the Lorentz group SO$(d,2)$ that rotates the $X^M_i$. 
The full bosonic
subgroup of $G_{L}$ may contain an additional factor $h_{L}.$ The matrices $%
\Gamma _{MN}$ (up to fixed coupling coefficients that depend only on $d,k$)
represent SO$\left( d-k,2\right) \times $SO$\left( k\right) $ and provide
the coupling of $L^{MN}$ to the appropriate part of the Cartan connection.
If there is an additional bosonic subgroup $h_{L}$ then the $L^{MN}$ have
zero coupling to the corresponding part of the Cartan form.

For the right
choices of $G$ the fermions $\Theta $ transform as a spinor at least under
SO$\left( d-k,2\right) $ (and either spinor or other representation under SO$%
\left( k\right) \times h_{L}$). Then they have the correct \textit{spacetime
spinor properties} to become the supercoordinates for a spacetime with
several supersymmetries. The structure of the Lagrangian above guarantees
the following local and global symmetries.

\begin{itemize}
\item  There is a local Sp$\left( 2,R\right) $ symmetry. The first term in (%
\ref{Lsuper}) is invariant just like (\ref{L}). The second term is invariant
because $L^{MN}$ is invariant and $g$ is a singlet under Sp$\left(
2,R\right) .$

\item  There is local symmetry under the bosonic subgroup SO$\left(
d-k,2\right) \times $SO$\left( k\right) \times h_{L}\in G_{L}.$ The SO$%
\left( d-k,2\right) \times $SO$\left( k\right) $ part also transforms $%
X_{i}^{M}$ locally. This can be used to eliminate part or all of the bosonic
degrees of freedom in $g,$ or part of the $X_{i}^{M}.$ In this way one may
rewrite the theory either in terms of spacetime vectors or spacetime
twistors (see \cite{twistor2T} for an example).

\item  There is a kappa-type local supersymmetry under part or all of the
fermionic transformations in $G_{L}.$ The local parameters kappa are
isomorphic to $\Theta $ with some projectors applied to it. The projectors
take into account the constraints $X_{i}\cdot X_{j}\sim 0$ that arise
because of the local Sp$\left( 2,R\right) .$ Because of the projectors kappa
supersymmetry can remove part but not all of the $\Theta \left( \tau \right) 
$ degrees of freedom. The fraction of $\Theta $ that can be removed ranges
from 0 to 3/4 depending on the specific details of the couplings in (\ref
{Lsuper}).

\item  There is a global supersymmetry $G_{R}$ under \textit{right
multiplication} of $g\rightarrow gg_{R}$. The $X_{i}^{M}$ do not transform
when all gauge degrees of freedom in $g$ are included. This is the global
spacetime supersymmetry $G_{R}$. It mixes only the bosons and fermions in $%
g, $ and it is realized \textit{linearly} on them. Although the $X_{i}^{M}$
do not transform in this version, they do transform after eliminating some
bosons from $g$ by gauge fixing the SO$\left( d-k,2\right) \times $SO$\left(
k\right) \times h_{L}$ gauge symmetries. When $g$ is gauge
fixed, both $G_{L}$ and $G_{R}$ must act on it to maintain the gauge. Then the
global $G_{R}$ transformation is induced on the $X_{i}^{M}$ as a field
dependent (including $\Theta $ dependent) local SO$\left( d-k,2\right) \times $
SO$\left( k\right) $ transformation. The resulting transformation becomes
precisely the needed global spacetime supersymmetry in $d+2$ dimensions. In
this version, $G_{R}$ mixes $\Theta ,X_{i}^{M}$ (and any bosons in $g$
that may have remained after fixing the $G_{L}$ gauge symmetries). When the
2T theory is further fixed to some kappa gauge and some 1T \ Sp$\left(
2,R\right) $ gauge, the $d+2$ supersymmetry (that has twice as many
fermionic parameters) reduces to the correct spacetime supersymmetry in $d$
dimensions. It must be emphasized that since there are many possible 1T
gauges as in Fig.1, the resulting supersymmetry is the correct one for those
1T theories.
\end{itemize}

A few applications of the general scheme are given in the next three
sub-sections.

\subsection{superparticle in d=3,4,6}

The simplest example of a supersymmetric action was discussed in \cite
{super2t}\cite{twistor2T} for $d=3,4,6$ dimensions with $N$ supersymmetries.
For these dimensions we take $G=$OSp$\left( N|4\right) ,$ SU$\left(
2,2|N\right) ,$ OSp$\left( 8^{\ast }|N\right) $ respectively. These
supergroups contain the SO$\left( d,2\right) $ spacetime subgroups SO$\left(
3,2\right)$ $=$ Sp$\left( 4\right) $, SO$\left( 4,2\right) =$SU$\left(
2,2\right) ,$ SO$\left( 6,2\right) =$Spin$\left( 8^{\ast }\right) $
respectively to which $L$ is coupled (i.e. $k=0$). There is zero coupling to
the internal subgroups $h_{L}=$SO$\left( N\right) ,$ SU$\left( N\right) ,$ Sp%
$\left( N\right) $ respectively. The fermionic parameters or supercharges
are classified as $\left( 4,N\right) ,$ $\left( 4,N\right) +\left( \bar{4},%
\bar{N}\right) ,$ $\left( 8^{\ast },N\right) $ respectively where the $4,4+%
\bar{4},8^{\ast }$ correspond to the spinor representations of the
corresponding SO$\left( d,2\right) .$ When all the gauge fixing is done, in
the 1T gauge that corresponds to the massless particle of Fig.1, the action (%
\ref{Lsuper}) reduces to the following form in $d$ dimensional superspace ($%
x^{\mu },\theta _{\alpha }^{a})$ with $N$ supersymmetries $a=1,2,\cdots N$%
\begin{equation}
\pounds =\dot{x}\cdot p-\frac{1}{2}A^{22}p^{2}+\tilde{\theta}_{a}\gamma
\cdot {p}\partial _{\tau }\theta ^{a}\Rightarrow \frac{1}{2A^{22}}\left( 
\dot{x}^{\mu }+\tilde{\theta}_{a}\gamma ^{\mu }\partial _{\tau }\theta
^{a}\right) ^{2}.
\end{equation}
This is the well known action for the massless superparticle. In the special
dimensions $d=3,4,6$ it does indeed have the larger global supergroup
symmetry $G=$OSp$\left( N|4\right) ,$ SU$\left( 2,2|N\right) ,$ OSp$\left(
8^{\ast }|N\right) $ respectively that includes the additional $N$ hidden
superconformal symmetries \cite{schwarz}\cite{super2t}.

It is important to emphasize again that one could choose gauges that
correspond to other 1T particle systems such as those of Fig.1. One would
then find the supersymmetrized versions for all of them in the special
dimensions $d=3,4,6.$ These are new supersymmetric actions that were not
noticed before except for the case of the supersymmetric H-atom in $d=4$ 
\cite{dHoker}, but even then the larger hidden symmetries were not known.

\subsection{AdS$_{5}\times $ S$^{5}$ supersymmetric Kaluza-Klein Tower}

In the previous paragraph the spacetime and internal subgroups of OSp$\left(
8|4\right) ,$ SU$\left( 2,2|4\right) ,$ OSp$\left( 8^{\ast }|4\right) $ were
treated in an asymmetric manner. The result was a superparticle moving in
flat space in $d=3,4,6$ only. Since these supergroups describe the
supersymmetries in the curved spaces 
AdS$_{7}\times S_{4},$ AdS$_{5}\times S_{5},$ AdS$_{4}\times S_{7},$ 
respectively, one may wonder if there is a more
interesting treatment of the subgroups that would apply to these cases.

We describe here the AdS$_{5}\times S_{5}$ case discussed in \cite{tower}.
We take $g\in $ SU$\left( 2,2|4\right) $ and twelve dimensions $X^{M},P^{M}$
with signature $\left( 10,2\right) .$ We divide them into two sets of six
dimensions each, $X^{m},P^{m}$ with signature $\left( 4,2\right) $ and $%
X^{a},P^{a}$ with signature $\left( 6,0\right) .$ The SO$\left( 10,2\right) $
orbital angular momentum now has components $L^{mn},L^{ma},L^{ab}.$ In the
coupling scheme only $L^{mn}\sim $SU$\left( 2,2\right) ,$ and $L^{ab}\sim $SU%
$\left( 4\right) $ are coupled in the action (\ref{Lsuper}), while $h_{L}=0$
does not exist. The kappa supersymmetry has a parameter of the form $\xi
=L^{ma}\left( \Gamma _{m}\kappa \Gamma _{a}\right) $ where $\xi $ is an
infinitesimal group parameter in $G_{L}$ classified as $\left( 4,\bar{4}%
\right) $ under SU$\left( 2,2\right) \times $SU$\left( 4\right) \subset $ $%
G_{L}.$ The coset orbital angular momentum $L^{ma}=X^{m}P^{a}-X^{a}P^{m}$
plays the role of a projector applied on the free local fermionic parameter $%
\kappa \left( \tau \right) ,$ which is also classified as $\left( 4,\bar{4}%
\right) .$ The kappa transformation (a fermionic $G_{L}$ transformation
applied only $g$ and on $A^{ij}$) gives $\delta \pounds =\left( \delta
A^{ij}+K^{ij}\right) X_{i}\cdot X_{j}$ where $K^{ij}$ comes from the second
term in the action; it is a complicated expression that depends on $%
g,X_{i}^{M},\kappa .$ Note the important fact that $K^{ij}$ multiplies $%
X_{i}\cdot X_{j}$ involving both $X_{i}^{m}$ and $X_{i}^{a}$ in the
precise combination that corresponds to the constraint $X_i\cdot X_j$.
Then we can choose $\delta A^{ij}$ so that 
 $\delta A^{ij}+K^{ij}=0$ to have the kappa symmetry, $\delta \pounds =0$.

A possible gauge choice is the following. Using the Sp$\left( 2,R\right) $
local symmetry we can choose two gauges: the component $P^{+^{\prime }}$ of
the SO$\left( 4,2\right) $ vector $X^m$ vanishes for all $\tau ,$ 
$P^{+^{\prime }}=0,$ 
and the magnitude of the SO$\left( 6\right) $ vector is a $\tau $
independent constant $\left| X^{a}\right| =a$ . Thus the AdS$_{5}\times $S$%
^{5}$ curved background is created from the flat 12-dimensional background
with this gauge choice (details in \cite{lifting}). Furthermore, using the
local $G_{L}$ symmetry we can eliminate all the bosons and half of the
fermions from $g$. Then $g$ contains only 8 complex fermions or 16 real
fermions that are non-linearly coupled to the orbital AdS$_{5}\times $S$^{5} 
$ symmetry operators $L^{mn},L^{ab}$ (which themselves are now non-linearly
realized as given in \cite{lifting}).

The gauge fixed action describes the entire AdS$_{5}\times $S$^{5}$
supersymmetric Kaluza-Klein tower \cite{tower} which was previously derived
in the context of compactified 10D type IIB supergravity. Notice the
following simple facts. First, the 16 fermions provide Ramond type vacua
with 128 bosons and 128 fermions. These correspond just to the fields of
supergravity that can be reclassified under the global symmetry group SU$%
\left( 2,2|4\right) .$ They are now propagating in the AdS$_{5}\times $S$^{5}
$ background. Second, although the constraints $X^{2}=X\cdot P=0$ have been
explicitly solved in the chosen Sp$\left( 2,R\right) $ gauge, the remaining
12-dimensional constraint $P^{2}=0$ takes the following sketchy form 
\begin{equation}
Casimir\,\,of\,\,SO\left( 4,2\right) +Casimir\,\,of\,\,SO\left( 6\right) =0.
\end{equation}
Thus, the mass of the field as given by the Casimir\thinspace \thinspace
of\thinspace \thinspace SO$\left( 4,2\right) $ is now fully determined by
the Casimir\thinspace \thinspace of\thinspace \thinspace SO$\left( 6\right) .
$ The Casimir\thinspace \thinspace of\thinspace \thinspace SO$\left(
6\right) $ for the graviton tower is determined by the SO$\left( 6\right) $
representations that can be constructed from the traceless symmetrized
products of $X^{a}$. The traceless tensor with $l$ indices has the SO$\left(
6\right) $ Casimir eigenvalue $l\left( l+4\right) .$ The mass of the rest of
the supermultiplet is determined by the SU$\left( 2,2|4\right) $
supersymmetry, which is built in from the beginning.

\subsection{Toy M-model}

The toy ``M-model'' sketched here was introduced in \cite{liftM} and
analyzed in \cite{toyM}. It illustrates some of the general aspects of
M-theory in the context of 2T physics that is explained in the next section.
Consider the Lagrangian given by (\ref{Lsuper}) with $d+2=13,$ and $g\in $OSp%
$\left( 1|64\right) .$ The coupling $L$ is to the SO$\left( 11,2\right) $
subgroup of OSp$\left( 1|64\right) .$ It can be shown that in one gauge (the
one closest to the massless particle in Fig.1) the model describes $d=11$
particle super-coordinates $x^{\mu },\theta ,$ together with collective
coordinates $x_{\mu \nu },$ $x_{\mu _{1}\cdots \mu _{5}}$ for the 11D
2-brane and 5-brane. In this gauge the linearly realized part of OSp$\left(
1|64\right) $ is precisely the 11D extended superalgebra of eq.(\ref
{M-algebra}) with certain constraints among the commuting brane charges. The
model has enough gauge symmetries (including bosonic extensions of kappa
type supersymmetries) to remove all ghosts associated with all 11D-covariant
super degrees of freedom that occur in the 2-brane and 5-brane collective
degrees of freedom \cite{toyM}. The quantum toy ``M-model'' is unitary.
Furthermore, in this gauge all degrees of freedom can be recast to the
language of super twistors (using the techniques of \cite{twistor2T}). The
resulting formalism is similar to the super oscillators as in \cite
{barsgunaydin} with 16-``colors'', and we must also require
``color''-singlet physical states. The physical spectrum can then be easily
determined: 128-bosons and 128-fermions (states of 11D supergravity) in the
presence of 2-brane and 5-brane charges with certain relations among them.
The model in this gauge realizes a BPS representation (a short
representation) of OSp$(1|64)$, with 32 vanishing supercharges. The other
non-vanishing 32 supercharges correspond to 16 linearly realized
supersymmetries plus 16 non-linearly realized superconformal symmetries.

\section{M-theory and 2T}

Since we have demonstrated that well known physics, including the Standard
Model, Gravity, and possibly string theory (see next section), have a 2T
physics formulation which sheds a deeper light into symmetries and higher
dimensions, we are encouraged to apply the same concepts to M-theory. After
all, 2T physics started with hints in the mathematical structure of
M-theory, F-theory and S-theory as described in the introduction. In the
following we will refer to the underlying theory as ``M-theory'', but it
will be evident that the remarks can apply to M-theory, F-theory or S-theory
seen as corners of the underlying ``M-theory''.

For ``M-theory'' 2T physics makes a general prediction independent of any
details \cite{liftM}. It requires that the explicit and hidden global
symmetries of ``M-theory'' must be described by the non-Abelian OSp$\left(
1|64\right) $ superalgebra. This should be useful for the eventual
formulation of the fundamental theory. The reasoning is simple. It is likely
that hidden spacetime symmetries of ``M-theory'' would be displayed by
adding $\left( 1,1\right) $ dimensions to the 
11 dimensions already noticed, with signature (10,1). Then the 2T physics
version of ``M-theory'' would be based on 13 dimensions with (11,2)
signature. If one adds supersymmetry and requires SO$\left( 11,2\right) $
covariance, one must have SO$\left( 11,2\right) $ spinors that have 64
components. The unique closure of the 64 supercharges that contains SO$%
\left( 11,2\right) $ is OSp$(1|64)$. This then must be the global
supersymmetry of ``M-theory''. Note that we insist on the Non-Abelian
supergroup, not a contracted version. As shown in \cite{liftM}, the usual
11D superalgebra (\ref{M-algebra}) with commuting momenta and commuting
2-brane and 5-brane charges is part of the non-Abelian OSp$\left(
1|64\right) .$ The Abelian 11D M-superalgebra has a ``triangular" type
of embedding in the non-Abelian 13D superalgebra.

The 2T ``M-theory'' must also have a gauge symmetry analogous to Sp$%
\left( 2,R\right) $ (or its generalizations discussed in the previous
sections). Although the gauge symmetry is unknown at this time, 
its presence would eliminate ghosts on the one hand, and provide
1T ``holographic" pictures of the 13D theory in lower dimensions 
on other hand. Then we may
view various corners of M-theory as gauge fixed versions of
the 2T formulation to various 1T formulations. In that case we would expect
that every 1T version should have the OSp$\left( 1|64\right) $ symmetry,
with part of it linearly realized and the remainder non-linearly realized
and somewhat hidden.

Further support for this point of view can be found in the fact that OSp$%
(1|64)$ correctly contains the various supersymmetries of several corners of
S-theory, F-theory and M-theory, including 13D, 12D, 11D, 10D type-IIA, 10D
type-IIB, heterotic, type-I, and AdSxS supersymmetries. This has been
demonstrated in \cite{liftM}. The relevant non-Abelian AdS$\times $S
superalgebras can be fitted in OSp$\left( 1|64\right) $ only after some
contractions which could be associated with large $N$ limits that occur in
discussions involving the AdS-CFT correspondence. The uncontracted versions
of the relevant AdSxS supersymmetries contain additional brane charges such
that the closure is uniquely OSp$\left( 1|32\right) $ contained in OSp$%
\left( 1|64\right) $ \cite{liftM}\cite{ferr}. One can also include OSp$\left(
8^{\ast }|8\right) $ discussed recently \cite{hull8} in the list of
interesting algebras contained in OSp$\left( 1|64\right) $ (with some
contraction) as being relevant \ for describing some corners of ``M-theory''.

\section{Strings and Branes}

For a p-brane $X^{M}\left( \tau ,\sigma _{1},\cdots ,\sigma _{p}\right) $
the natural candidate for a local symmetry that replaces Sp$\left(
2,R\right) $ is the conformal group on the worldvolume SO$(p+1,2)$. This was
pointed out in \cite{conf}. Localizing this group implies coupling $%
X^{M}\left( \tau ,\vec{\sigma}\right) $ to the gauge fields of conformal
gravity on the brane. In particular for $p=0$ this yields the same
Lagrangian as local Sp$(2,R)$=SO$(1,2)$ in the second order formalism
obtained by integrating out $P^{M}$ in eq.(\ref{L}). A first order
formulation of this theory analogous to the Sp$\left( 2,R\right) $
formulation has been constructed \cite{ib2tstring}. The global symmetry is
again SO$\left( d,2\right) $ in flat space. The gauge symmetry and
constraints are such that $d+2$ is reduced to $d$ dimensions for the entire
brane for any $p$. A (partial ?) formulation of 2T physics for p-branes
based on this idea has been published in \cite{string2t} although in that paper
only a related and simplified version of \ the formalism was analyzed which
appeared useful at the time. It was shown that 1T tensionless and rigid
branes on flat or AdS$\times $S backgrounds could be obtained from the 2T
p-brane by choosing certain 1T gauges for SO$(p+1,2)$. A gauge that
describes branes with tension (like \textit{massive} particles of Fig.1)
seemed difficult to find. It was not understood whether the search for gauge
choices was not sufficiently broad or whether the Lagrangian could be
missing some terms. Thus, more work needs to be done to establish the 2T
formalism more generally for p-branes.

\section{Outlook}

The 2T point of view of ordinary 1T physics has revealed a new subtle
dimensional unification in $d+2$ dimensions. The 2T language is the only
explanation of the symmetries and the relations among the 1T dynamical
systems in the same family. Furthermore, the action looks simpler, more
elegant, more symmetric and displays a fundamental gauge principle which is
not apparent in 1T physics.

The gauge symmetry is similar to ``duality'' if analyzed from the point of
view of 1T systems. Each 1T system can be considered as a $d$ dimensional
``holographic'' image of the $d+2$ theory. Only, holography here gives 
an image projected on a ``surface" of two less spacetime dimensions. 
Each 1T holographic image appears as a different
dynamical system from the point of view of 1T physics. The global symmetry SO%
$\left( d,2\right) $ is observable in every holographic image. It is one of
the key evidences for 2T physics which could be verified experimentally or
computationaly either within a given 1T system (e.g. H-atom as an SO(4,2)
system, etc.) or in duality relations among several systems as outlined in a
previous section. In principle there are an infinite number of such duality
relations, and it would be an interesting project to compile some
predictions that could be tested experimentally or computationaly when such
computations are possible.

The known physics of the Standard Model has a 2T field theory description.
It appears as a 4-dimensional holographic image of 
a 2T theory in $4+2$ dimensions. The
general 2T field theory formulation can probably be improved and put into a
more fundamental form so that all equations (kinematic + dynamic) follow
from the same field theory. This more general formulation may involve
aspects of non-commutative geometry.

Based on the insights brought by 2T physics into 1T physics, it is
reasonable to expect that ``M-theory'' has \ sufficient hidden symmetries
that amount to as much as a non-Abelian OSp$\left( 1|64\right) .$ It may be
helpful to use this remark as a guide to build more systematics about
``M-theory'' and to come closer to an eventual construction of the theory.

There are many doable problems in furthering the techniques of 2T physics,
using it in new physical applications or computations involving ordinary
macroscopic or microscopic physics, and finding experimental tests to verify
some of its predictions. There are also some technical and conceptual
projects that include the following

\begin{itemize}
\item  Completion of the 2T formalism for strings and branes.

\item  Analysis of various aspects of 2T field theory, including relations
among 1T field theories, in particular those related to QCD or the Standard
Model. Can we learn something new and non-perturbative about QCD or the
Standard Model in this process?

\item  Backgrounds in the presence of supersymmetry. This is expected to
lead to supersymmetric Yang-Mills and supergravity in $d+2$ dimensions. In
particular, we know that backgrounds for the 10D superparticle must obey the
super Yang-Mills equations to preserve the kappa supersymmetry. The 2T
formulation of this system must then yield (10,2) super Yang-Mills including
the dynamical equations, not only the kinematic ones. The same is expected
about supergravity in (10,2) or (11,2) dimensions. These are theories that
have been sought for but never constructed satisfactorily before. The 2T
approach might do it.

\item  There is renewed interest in higher spin fields. These have been
mainly analyzed in $d=3$ dimensions based on the conformal group SO$\left(
3,2\right) $ $=$ $Sp\left( 4\right) $ \cite{vasil2}\cite{sezgsundr}. The
background field scheme involving higher powers of $P^{M}$ described in a
previous section provides a new formalism for discovering the properties and
the field equations for higher spin fields in an explicitly SO$\left(
d,2\right) $ covariant approach in any dimension.

\item  Construction of ``M-theory''.
\end{itemize}

What message is there about spacetime? Are there really two times, are there
really 13 dimensions? The formalism of 2T physics teaches us that one extra
spacelike and one extra timelike dimensions do exist, but not in the naive
sense. The usual $\ 4$ (or $d$) dimensions together with the extra 2
dimensions define a particular ``holographic'' view of a 2T system in a
spacetime with 6 (or $d+2$) dimensions. This ``holographic'' picture appears
as a dynamical system described by 1T physics in 4 (or $d$) dimensions. A
different 1T dynamical system is obtained from a different ``holographic''
view of the same 2T system by embedding 4 (or $d$) plus 2 dimensions
differently in 6 (or $d+2$) dimensions. To obtain all possible
``holographic'' views, corresponding to all possible 1T dynamical systems in
the same family, all the $d+2$ dimensions are needed on an equivalent basis.
In this sense all the $d+2$ dimensions exist. They reveal themselves in any
of the holographic pictures described by 1T physics by the presence of
hidden symmetries and in the relations among many 1T systems. 2T physics
teaches us that spacetime can be more than what we used to think of before.

\end{document}